# Quantum Dot Moiré from Crossed $MoS_2$ Nanoribbons

Xinting Shuai[1], Hao Zhang[3,4], Wenjing Wu[2,4], Chongning Wu[1], Maryam Amiri[3], T. A. M. Ragib Shahriar[1,4], Dian Pan[1,3], Zhi Kai Ng[1,5], Tymofii Pieshkov[1,4], Leeza Dutta[1,6,7], Yijun Zhou[1], Rohith Narra[3,4], Luke Van Leeuwen[1], Jishnu Murukeshan[1], Luyao Shi[1], Jiawei Lai[1], Atin Pramanik[1], Bipin Kumar Gupta[6,7], Edwin Hang Tong Teo[5], Robert Vajtai[1], Xiang Zhang[1], Hanyu Zhu[1,2], Shengxi Huang[2], Aditya D. Mohite[3], Pulickel M. Ajayan[1]*

## Affiliations:

[1] Department of Materials Science and NanoEngineering, Rice University, Houston, Texas, 77005, United States

[2] Department of Electrical and Computer Engineering and the Rice Advanced Materials Institute, Rice University, Houston, Texas, 77005, United States

[3] Department of Chemical and Biomolecular Engineering, Rice University, Houston, Texas, 77005, United States

[4] Applied Physics Graduate Program, Smalley-Curl Institution, Rice University, Houston, TX 77005, United States

[5] School of Electrical and Electronic Engineering, Nanyang Technological University, Singapore, 639798, Singapore

[6] Photonic Materials Metrology Sub Division, Advanced Materials and Device Metrology Division, CSIR-National Physical Laboratory, New Delhi-110012, India

[7] Academy of Scientific and Innovative Research (AcSIR), Ghaziabad-201002, India

**Corresponding author. Email: ajayan@rice.edu**

**Abstract**: Twisted atomically thin layers have attracted much attention for Moiré potential and correlated quantum phenomena. However, existing Moiré superlattices have largely been limited to extensive wavefunction without lateral confinement. Here we introduce a new platform where 1D nanoribbons of 2D $MoS_2$ grown by vapor deposition can be easily superposed at various angles from stacking and transferring, to form Moiré quantum dots at their intersections with unique exciton physics. Angle-dependent Moiré intersections show enhanced exciton emission at commensurate angle 22°, which demonstrates faster relaxation at the cryogenic temperature. A size-dependent study further exhibits a reduced exciton energy and soften out-of-plane interlayer coupling for smaller Moirés areas. Our results reveal exciton physics turnability via precise overlapping of 1D nanoribbons.

## Main

Moiré superlattices formed by twisting crystals stacked provide a versatile route to engineer quantum states at their van der Waals (vdW) interface. By varying the twist angle, the spatially periodic potential, or the Moiré potential changes, reshaping electronic bands and excitonic selection rules. This has since resulted in groundbreaking fundamental discoveries, such as the observation of Hofstadter butterfly and Brown–Zak oscillations.[1] In semiconducting TMDs, Moiré potentials can localize excitons into deep, registry-defined minima that behave as ordered arrays of quantum-dot-like sites, enabling angle-tunable emission and presents new opportunities for quantum optoelectronics.[2–4] More broadly, because the Moiré potential is tied to local stacking symmetry, it can also influence valley-contrasting optical selection rules and related spin–valley responses.[5,6]

To date, Moiré excitonics have been explored predominantly in 2D-2D twisted bilayers, where interlayer rotation generates extended Moiré superlattices and has revealed rich photo-physics in a global scale for the interlayer Morie exciton properties such as Moiré trapped excitons and twist angle tunable optical signatures.[7–9] In these systems, lattice relaxation, reconstruction, are now recognized as important ingredients that shape the Moiré landscape and enrich the optical response.[10,11] At low temperature, the observed emission can also be shaped by competition between bright channels and spin- or momentum- forbidden dark states, motivating careful control of local structure and geometry.[12,13] However, a key challenge for extended 2D-2D Moiré superlattices is that, beyond twist-angle control, they offer limited intrinsic lateral confinement and device-to-device geometric tunability, and their measured response is often complicated by spatial variations in twist angle and heterostrain during fabrication or annealing, which can produce non-uniform Moiré potentials and interface cleanliness issues. By contrast, Moiré formation in laterally confined 1D building blocks, such as crossed nanoribbons (NRs), offers a complementary regime in which the Moiré region is restricted to a small overlap junction and boundary/edge effects can become more pronounced, with geometric confinement expected to influence exciton capture and relaxation.[14,15] However, this micro scale/confined 1D-1D Moiré system remains largely unexplored.

Here, utilizing catalyst-free epitaxial chemical vapor deposition CVD synthesis of monolayer $MoS_2$ NRs on mica, followed by deterministic stacking with controlled twist angle, we developed a Moiré platform based on twisting 1D $MoS_2$ NRs to study such a phenomenon. The width of the NRs confines the overlap region into a nanoscale junction that resembles more of a quasi-0D quantum-dot system rather than an extended 2D superlattice, and hence we name such a system Moiré quantum-dots (MQDs). We observed remarkable phenomenon that were not explainable by 2D Moiré physics, such as the stronger interlayer interactions over a wide range of twist angles, and a counterintuitive increase in emission intensity of a stacked $MoS_2$ MQDs near commensurate angle $\theta \approx 22°$. Using temperature dependent Photoluminescence (PL) we also observed an increased emission and localized fast radiative channel at cryogenic temperature. Size-dependent analysis further demonstrated a PL red-shift in smaller MQDs, and edge-mediated lattice reconstruction deepens the local Moiré potential. This platform provides a new approach to studying commensurate angle Moiré physics at quantum dot scale and can be extended to other pairs of 1D materials.

## Assembly of a 1D-1D Moiré platform

The purity of $MoS_2$ NRs is critical to the reliability of the subsequent experiments and Moiré observations. Hence, we developed a catalyst-free one-step CVD method, where $MoO_3$, NaCl and sulfur were used as precursors. Briefly, the freshly-cleaved mica was loaded on the top of the precursor boat downstream, and monolayer $MoS_2$ NRs would be formed on backside of the substrate (face-up) due to lowered density of $MoO_3$ vapors during the growth (See Supplementary Fig. 1a and section 1.1 for more information). From Fig. 1b, $MoS_2$ NRs horizontally grown on mica substrate have an angle of 60° or 120° between most pairs of interconnected NRs, following the orientation of single crystal mica (001) surface due to the CVD epitaxial growth, which is consistent with previous reports on epitaxial CVD growth of NR or nanowire nanomaterials on single crystal mica and c-sapphire substrates.[16–18] The synthesized NRs have width ranging from 30 nm to 1300 nm, with most occurrences from 50 nm to 400 nm (Supplementary Fig. 1e). The Raman spectrum of the $MoS_2$ NRs (Supplementary Fig. 1c) revealed the presence of characteristic phonon modes of $MoS_2$, with $E_{2g}^1$ and $A_{1g}$ peaks observed at 383.8 and 402.5 $cm^{-1}$, respectively. The energy gap between these two peaks was estimated to be 18.7 $cm^{-1}$, which is consistent with that of monolayer $MoS_2$.[19] Furthermore, strong PL intensity was observed at around 1.88 eV upon excitation with a 532 nm laser (Supplementary Fig. 1d), indicating the direct band-gap character of monolayer $MoS_2$.[20] From the AFM scanning image (Supplementary Fig. 1b), the height of the ribbon on mica substrate is around 0.75 nm, which is consistent with the monolayer $MoS_2$ thickness.[21] From SEM image of NRs transferred to $SiO_2$/Si substrate (Supplementary Fig. 2c), some NRs are terminated with a tip particle. Both AFM topography and phase images (Supplementary Fig. 2a) confirmed the tip particles at the end of some NRs, which confirms the formation of the monolayer $MoS_2$ NRs is through the Vapor-Liquid-Solid (VLS) process, where the liquid phase precursor droplets migrate laterally on the mica surface before precipitating.[22–24] The elemental mapping images and the spectrum (Supplementary Figs. 2c and d) confirmed the presence of only Mo and S and the absence of Na or Cl in the tip particle. The Raman peak (Supplementary Fig. 2b) confirmed the only $MoS_2$ characterization peaks in the particle without any $MoO_3$ or $MoO_2$ content. We also observed some NRs originated from large nucleation particles without terminated tips and the EDAX mapping (Supplementary Fig. 2e) shows the same element compounds of nucleation particle with terminated tip particles. This indicates three main steps during the growth: (1) Vapor precursors deposits onto the mica substrate and forms $MoS_2$ particles immediately at the nucleation site; (2) The $MoS_2$ paticles are partially turned into $MoS_2$ liquid droplets; (3) The solid state of monolayer NRs can be epitaxially formed through the consumption process of $MoS_2$ droplets, if the droplets is fully consumed during growth, there would be no tips remaining in the termination sites. To the best of our knowledge, this is the first reported successful synthesis of monolayer $MoS_2$ NRs with no assistance of salt or metal catalysts on sample.

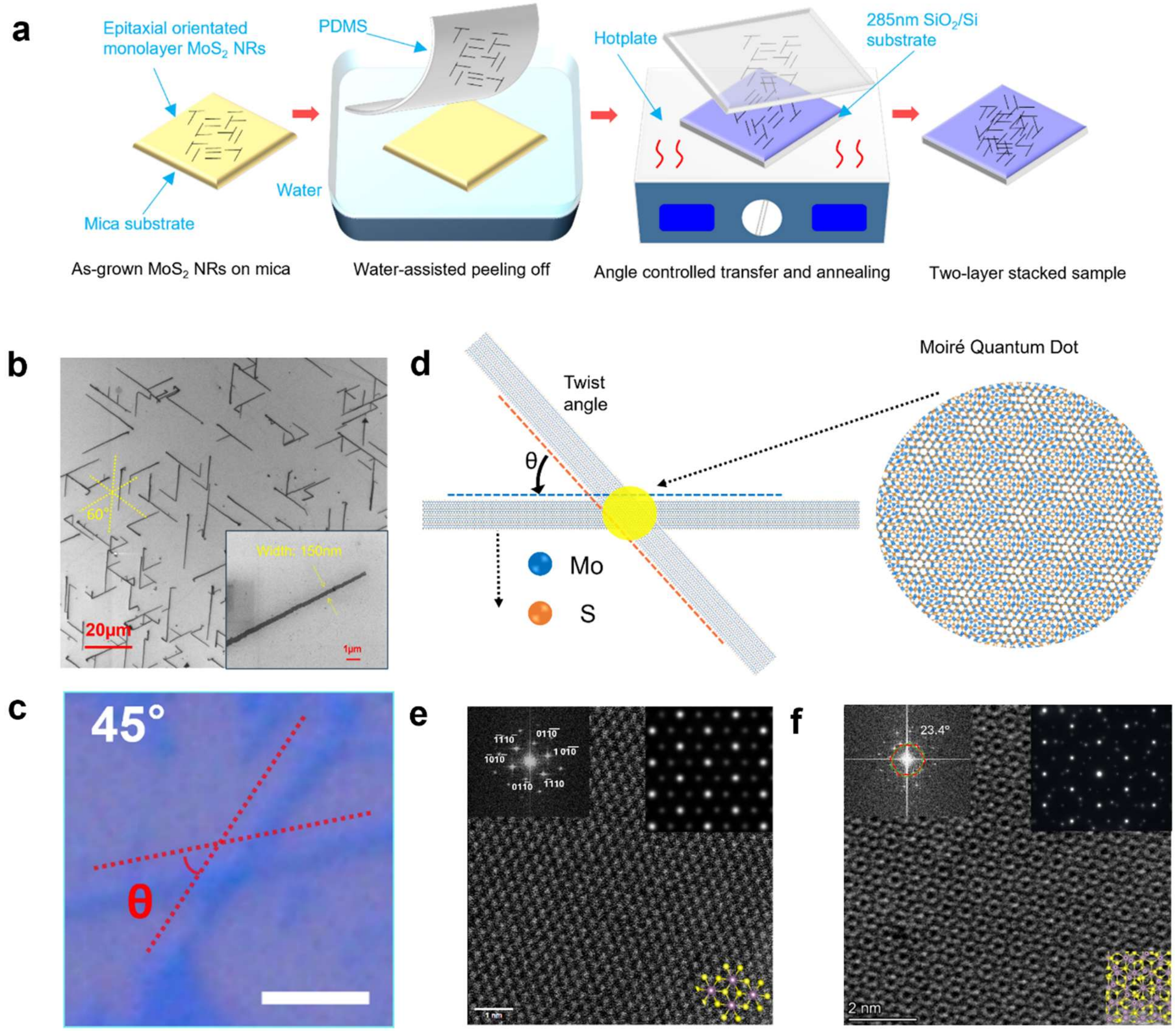

**Fig. 1: Fabrication of $MoS_2$ MQDs a,** Schematic of MQDs creation through $MoS_2$ NRs as-grown $MoS_2$ NRs on mica, water-assisted peeling off, angle-controlled transfer and annealing and two-layer sample stack. **b,** SEM image of $MoS_2$ grown on mica by CVD method, interconnected NRs have 60° or 120° angle. Inset is high magnification SEM image of single NR with 150nm width. **c,** Optical microscope (OM) image of MQD from two-layer NRs stacking and 45° twisting, scale bar is 2μm. **d,** Atomic schematic of (left) stacked $MoS_2$ NRs and MQD with controlled twist angle θ, and (right) zoomed in Moiré pattern from MQD, blue and orange atoms are Mo and S separately. **e,f,** STEM image of corresponding atomic configuration of monolayer NR (**e**) and MQD from bilayer stacking (**f**) with 23.4° twist angle, with the corresponding FFT (top-left inset), atomic-scale overlay (bottom-right inset) and the simulated diffraction patterns (top-right inset).

Unlike conventional 2D-2D bilayers, the NRs geometry inherently defines a 1D building block with a deterministic axis and a laterally confined overlap region. Since majority of the NRs' width ranges from 50 nm to 400 nm, we expect the MQDs formed by twisting and stacking the NRs to be approximately in the range as well. From Fig. 1a, as-grown epitaxially oriented NRs on mica can be peeled off by PDMS film due to vdW force between $MoS_2$ and mica substrate.[25] During the peeling off process, $MoS_2$ attached with PDMS was immersed into DI water to introduce water molecules naturally to wet the mica surface, the uniform $MoS_2$ NRs can be lifted off smoothly due to the hydrophobic and hydrophilic properties of 2D materials and mica substrate.[26] The first layer of $MoS_2$ NRs can be formed by transferring NRs to 285 nm $SiO_2$/Si substrate, the second layer of NRs on PDMS were stacked on the first layer via a transfer stage. Since the orientation of NRs on each layer is fixed, the twist angle can be controlled by rotation transfer stage. After annealing the sample on hotplate, the PDMS film is removed, and the stacked NRs remain on $SiO_2$/Si substrate (Fig. 1c), forming our quantum dots with Moiré pattern in Fig. 1d. This is also the first time forming a Moiré pattern on such a small scale.

To further investigate the MQDs in cross-section bilayer area, the scanning transmission electron microscopy (STEM) was conducted. Fig. 1e presents a high-resolution STEM image of a monolayer of $MoS_2$, exhibiting the characteristic hexagonal symmetry of the 2H phase. The corresponding fast Fourier transform (FFT) pattern (Figure 1e, top inset) reveals a single set of sharp, discrete reflections indexed along the [001] zone axis, confirming high crystallinity and phase purity. The superimposed atomic model further illustrates the trigonal prismatic coordination of Mo and S atoms, consistent with the 2H-$MoS_2$ polytype. In contrast, Fig. 1f presents a STEM image of a pronounced Moiré superlattice arising from the vertical stacking of two $MoS_2$ layers with a relative rotational misalignment. The real-space image displays a periodic interference pattern with an enlarged periodicity relative to the intrinsic lattice. The corresponding FFT (Fig. 1f, top-left insets) exhibits two distinct sets of hexagonal diffraction spots; the angular separation between these reciprocal lattices corresponds to a twist angle of approximately 23.4°. This rotational mismatch gives rise to long-range Moiré interference, further visualized in the atomic-scale overlay (bottom-right). To validate these structural assignments, simulations were performed using the abTEM framework (See section S1.9). The simulated diffraction patterns and image contrast (top-right insets) show excellent agreement with the experimental data, accurately reproducing both the monolayer symmetry and the complex Moiré interference features.

To further characterize $MoS_2$ MQDs, the PL spectra of the 35° MQD and connected monolayer single ribbon was shown in Fig. 2a, the two main peaks are located at 1.867 and 1.88 eV. We observed a 13 meV red-shift and lower PL intensity in MQD, which are attributed to the binding energy difference and carrier redistribution between monolayer and bilayer NRs.[27,28] The Raman spectra and inset OM image in Fig. 2b show larger characterization peaks difference and higher Raman intensity at 48° MQD compared with connected monolayer ribbon, which are consistent with previous reports.[29,30] Raman mapping (Supplementary Fig. 3a and b) showed the intensity spatial mapping of $E_{2g}^1$ mode in grey color and $A_{1g}$ mode in red color, both Raman mapping displayed the distinct peak difference and color contrast between stacked NRs and substrate, also suggest uniformity of the layers.

The Raman position mapping in Fig. 2c ranges from 373.5 to 391 $cm^{-1}$ indicated a clear contrast of $E_{2g}^{1}$ between the monolayer single ribbon and bilayer MQD, which located at 386 and 382 $cm^{-1}$, respectively. The other yellow dots on single ribbon might be from the nucleation sites.

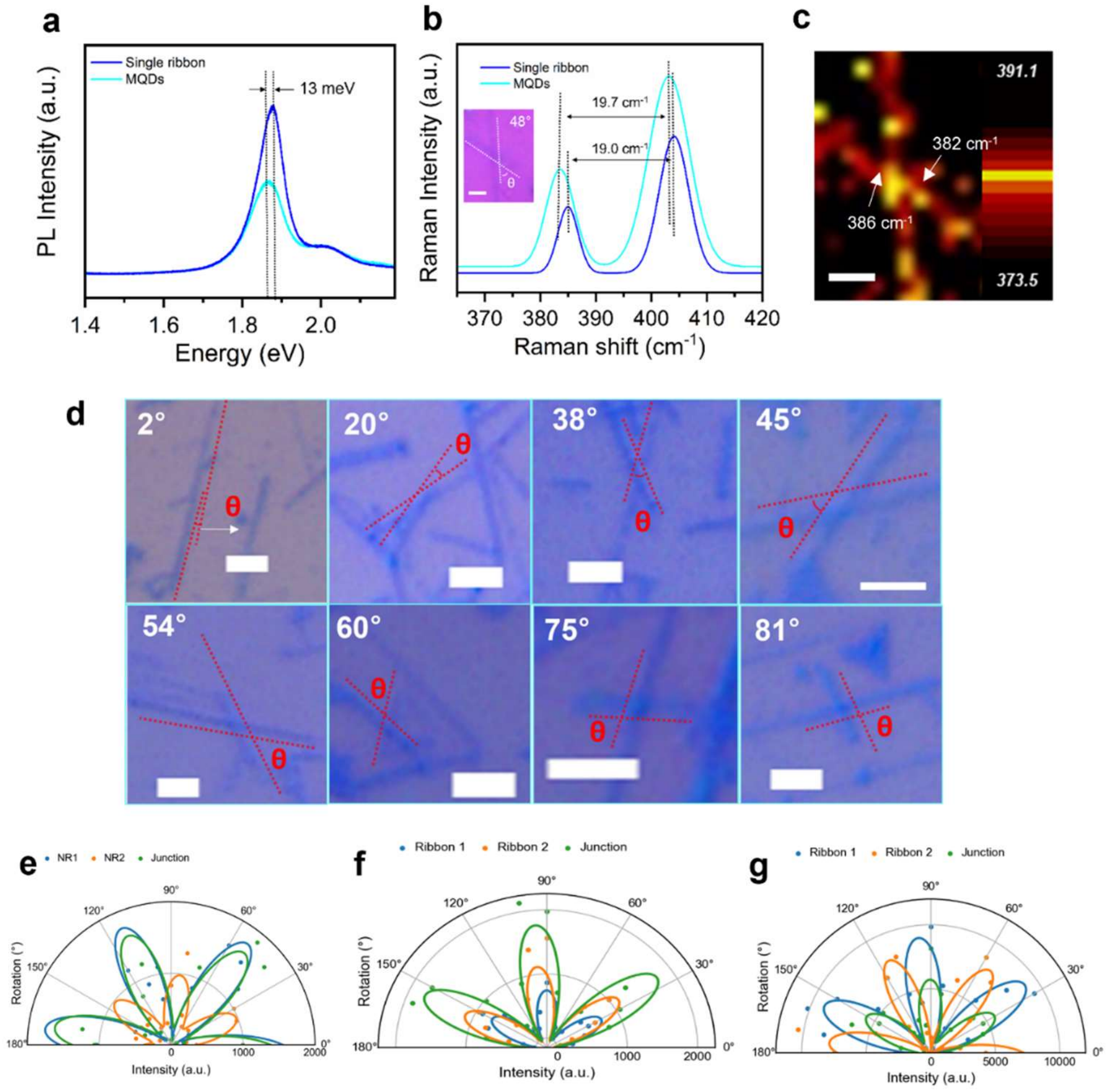

**Fig. 2: Characterization of $MoS_2$ MQDs a,b,** Typical PL spectra (**a**) of monolayer single NR and 35° MQDs and Raman spectra (**b**) of monolayer single NR and 48° MQDs, showing different PL and Raman peak positions between single NR and MQDs (OM image of 48°C MDQ inset, scale bar is 1 μm). **c,** Raman position mapping ranges from 373 to 391.1 $cm^{-1}$ with 48° MQDs, scale bar is 1μm. **d,** OM images of stacked MQDs from 0 to 90°, the red dashed line indicates the corresponding NR and twist angle, scale bar is 2μm. **e-g,** Polar plot of second harmonic generation (SHG) intensity of MQDs and NRs with 90° (**e**) and 53° (**f**) and 41° (**g**) twist angle; green, blue and orange colors indicate junction of MQDs, interconnect NR1 and NR2, respectively.

The OM images in Fig. 2d showed stacked MQDs with different twist angles, angles between the sharp edge cross-section of two layers in images are 2°, 20°, 38°, 45°, 54°, 60°, 75° and 81°, respectively, demonstrating the wide range of MQDs with twist angles from 0 to 90° can be made by our fabrication method. Confined overlap area also provides an independent geometric control parameter that 2D film-on-film lacks. To determine the crystallographic orientation of each $MoS_2$ NR and the twisting angle of the stacked NRs, polarization-resolved second-harmonic generation (SHG) was measured as a function of orientation for three spots, with two NRs at twist angles (θ) of 90°, 53° and 41° (Supplementary Figs. 4a-c). The incident and detected polarizations were configured in the parallel geometry, with the analyzer parallel to the incident polarization, yielding SHG maxima along the armchair direction of each $MoS_2$ NR.[31,32] The sample was rotated by the azimuthal angle φ about the surface normal. Owing to the non-centrosymmetric $D_{3h}^1$ structure of monolayer $MoS_2$,[32] each NR exhibits a pronounced six-fold polar pattern (Fig. 2e-g) characteristic of the underlying threefold lattice symmetry.[31] The single NR responses were fit to $I(\varphi) = A\cos^2(3(\varphi - \varphi_0)) + C$,[33] where $\varphi_0$ is the crystal orientation angle, A is the SHG amplitude, and C is the background, yielding each ribbon's orientation. Because $\cos^2(3\varphi)$ is 60°-periodic, the extracted single-NR orientation is defined modulo 60°. However, the relative twist between two NRs should be 120°- periodic, so there is still ambiguity between $\theta = \varphi_{0,NR1} - \varphi_{0,NR2}$ and $\theta = \varphi_{0,NR1} - \varphi_{0,NR2} + 60°$. To resolve this ambiguity, we analyzed the coherent SHG interference at the NR junction (see the full fitting procedure in Supplementary 1.1). The twist angles extracted from the junction fits are 30.2°, -6.4° and 41.8°41.8°, -6.4°, and 30.2° for the three samples (Supplementary Fig. 3a-c), respectively. Therefore, sample 3 retains its nominal assignment, but samples 1 and 2 are reassigned to $\theta_{OM} + 60°$ (or equivalently $60° - \theta_{OM}$ since we do not differentiate left- and right- handed twists within the scope of this work). The ability to differentiate between AA-like and AB-like stacking is important for assigning the observed properties of NR junctions to the correct angle within the 0°-60° interval.[33,34]

## Twist-angle dependent characterization of MQDs

Raman spectroscopy was used to quantify the twist-angle-dependent lattice response of $MoS_2$ NR bilayers. While in an extended 2D bilayer, θ and 120 + θ should be equivalent by symmetry.[35] In MQDs, however, the finite ribbon axes, edge orientations, and overlap shape are not generally equivalent under this operation. Even if the local Moiré period is the same, the finite-size Moiré truncation, edge contribution, and strain relaxation can be different. Considering the rotational symmetry and nanometer scale of monolayer $D_{3h}^1$ and bilayer $D_{3d}^3$ $MoS_2$ NR,[36] therefore the twist-angle series for Raman and PL was analyzed over the reduced interval of 0°- 60° (Supplementary Fig. 5a). Fig. 3a and extracted peak positions in Fig. 3b show the Raman spectra in the 370-415 $cm^{-1}$ range for representative twist angles, where the in-plane $E_{2g}^1$ and out-of-plane $A_{1g}$ modes are resolved. With increasing twist angle, the $E_{2g}^1$ mode exhibits a pronounced blue shift and reaches its highest frequency near ~30°. This evolution of the in-plane phonon is consistent with prior reports that interlayer coupling and equilibrium interlayer separation depend strongly on stacking registry, being weakest near ~30° and strongest near 0°/60° in twisted bilayer $MoS_2$ systems.[37,38] However, the $A_{1g}$ mode remains weakly changed, indicates stronger interactions among different twist angles than 2D Moiré superlattice.[36,39]

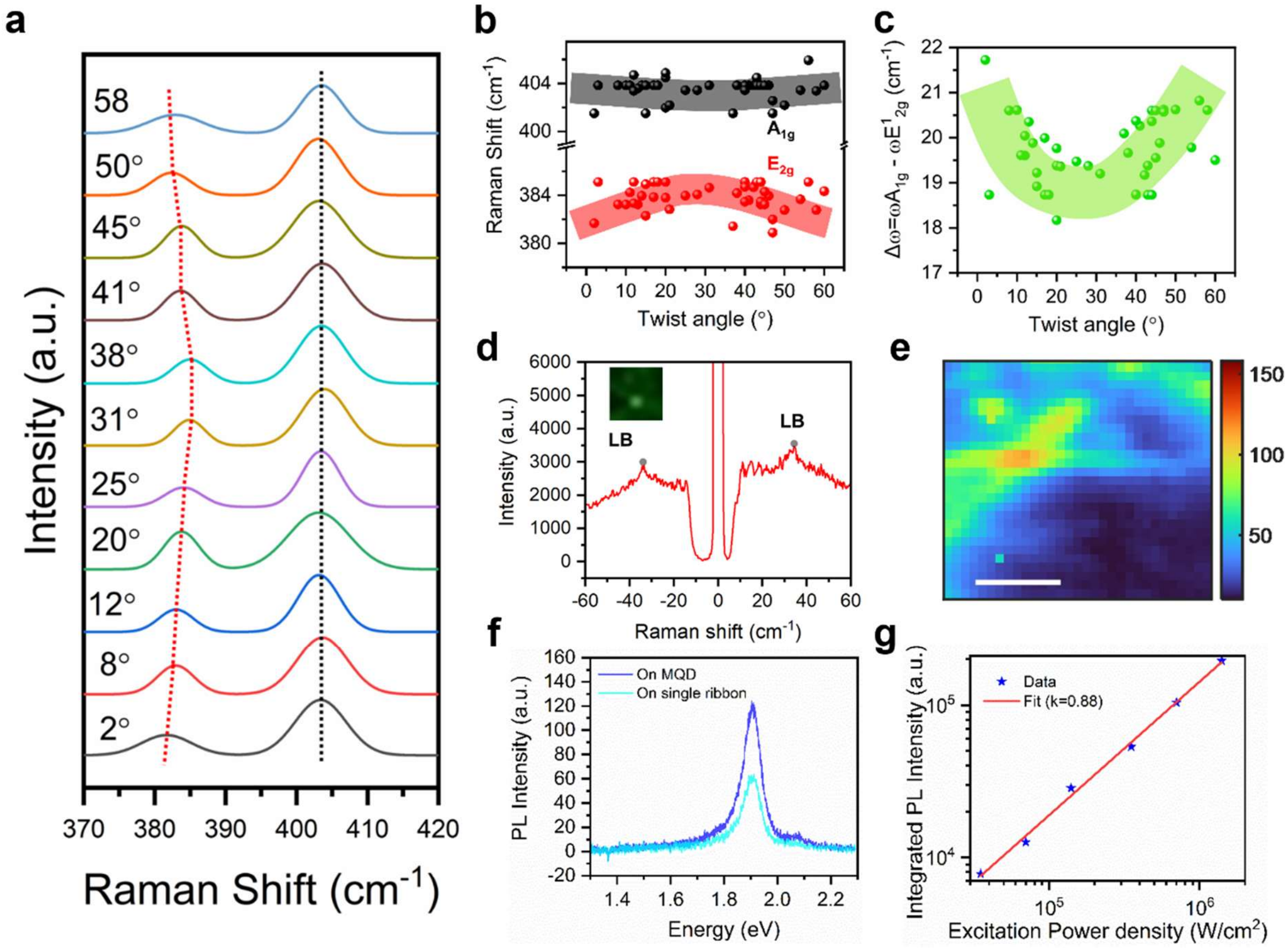

**Fig. 3: Twist angle-dependence for spectroscopy of the MQDs a-c,** Raman spectra (**a**), Raman position ($A_{1g}$ and $E^1_{2g}$) (**b**) and Raman frequency difference ($\omega A_{1g} - \omega E^1_{2g}$) (**c**) of the MQDs area as the dependence of twist angle from 0° to 60°, red and black dashed line in (**a**) indicate the $E^1_{2g}$ and $A_{1g}$ peaks change trend with different angles. **d,** Low-frequency Raman spectra of the MQDs area with the Raman mapping (frequency from 35-40 $cm^{-1}$) (inset). **e,f,** 7K low-temperature PL mapping (**e**) of the stacking $MoS_2$ NRs (scale bar is 1 μm) and spectrum comparison (**f**) of the 47° MQD and the interconnected single NR area. **g,** Experimental (blue star) and fitting curve (red line) of power dependence for 47° MQD area, the fitted slope is 0.88.

The twist-angle dependence of the mode separation, $\Delta\omega = \omega A_{1g} - \omega E_{2g}^{1}$, is summarized in Fig. 3c. Δω decreases as the twist angle approaches ~30° and exhibits a global minimum near this angle, consistent with the $E_{2g}^{1}$ hardening observed in Fig. 3a and b. The $A_{1g}$ - $E_{2g}^{1}$ separation is widely used as a convenient Raman metric for layer thickness and variations in interlayer interactions and stacking configuration in $MoS_2$ based layered structures.[40] Notably, in certain twist-angle MQDs area, Δω approaches ~20 $cm^{-1}$ or smaller, which is reduced relative to typical bilayer-like separations and suggests locally weakened coupling. Related reductions in effective interlayer coupling have been observed in twisted or folded $MoS_2$ structures and in artificially assembled stacks where interface cleanliness and nanoscale corrugation modulate the coupling.[41,42] Finally, the fact that both the peak positions and Δω vary across twist angles is consistent with twist-induced lattice relaxation producing spatially non-uniform strain and interlayer spacing, which are known to shift $MoS_2$ Raman modes.[43,44]

To further assess the interlayer coupling in the stacked $MoS_2$ NR bilayers, we performed low-frequency (LF) Raman spectroscopy in the $\pm 60$ $cm^{-1}$ window in Fig. 3d, where the interlayer shear (S) and layer-breathing (LB) phonons provide a more direct probe of the interfacial mechanical restoring forces than the intralayer $A_{1g}$ and $E_{2g}^{1}$modes.[45–47] At a representative twist angle of 25°, a clear LB mode at 39.08 $cm^{-1}$ is observed, and LF spectra acquired over the full twist-angle series (Supplementary Fig. 6a) consistently show the presence of the LB mode. Moreover, inserted LF Raman mapping of the LB mode at 58° reveals that the LB mode signal is confined to the NR overlap junction, directly identifying the bilayer region and indicating a finite out-of-plane interlayer restoring force. The shear mode is not resolved in the NR junction, which is consistent with the very limited overlap and with twisted or incommensurate bilayers where the shear response can be strongly suppressed.[48]

To study the photo physics of MQDs under different temperatures, the PL spectrum comparison between MQD cross-section and adjacent single ribbon under RT and LT is collected at a 47° twisted angle. At RT, the PL signal for single ribbon is slightly higher than MQD signal (Supplementary Fig. 7a). At low temperature (7 K), spatially resolved PL in Fig. 3e reveals that the emission is strongly enhanced at the NR overlap for a nominal 47° twist, with the PL intensity at the cross section approximately two-fold higher than that of the adjacent monolayer NRs in the PL spectrum of Fig. 3f. This behavior is counterintuitive because $MoS_2$ generally becomes less emissive as the layer number increases, consistent with the evolution from a direct-gap monolayer to an indirect-gap bilayer or multilayer electronic structure.[49–51] The enhanced PL at the junction suggests that the twist induced Moiré potential and local registry can modify intralayer exciton recombination at this angle,[5] effectively brightening localized excitons in the overlap region.[52] Consistent with this picture, the excitation power dependence extracted at the MQD region yields a sublinear exponent with 0.88 slope as shown in Fig. 3g, indicative of recombination dominated by localized states that approach saturation and are commonly associated with Moiré or strain modulated trapping potentials in twisted TMD stacks.[53–55] Taken together with the twist dependent Raman shifts discussed above, the enhanced and nonlinearly scaling PL at the MQD region supports a picture in which Moiré driven lattice relaxation and junction potential generate a spatially inhomogeneous excitonic landscape in the overlap, providing a natural starting point for analyzing twist tunable exciton dynamic in the following section.

## Excitonic photo-physics of MQDs

To further investigate the nature of exciton physics at the MQDs, we systematically analyzed the PL spectrum under RT and LT (T= 3 K) as a function of different twist angles. RT PL from 0 to 60° MQDs can be consistently deconvolved into the $X^A$ and $X^B$ excitons and the trion $X^{A-}$, with Lorentz mode fitted peak positions clustered at 1.86-1.88 eV ($X^A$), 2.00-2.03 eV ($X^B$) and 1.82-1.85 eV ($X^{A-}$) across the series (Fig. 4a and Supplementary Fig. 8a). Different from other 2D $MoS_2$/$MoS_2$ twisted homostructure literatures,[34,52] we didn't observe any low energy indirect bandgap Moiré exciton located around 1.6 eV under RT or LT, which might have resulted from low PL signal from our sub-micron MQDs scale. We quantified and compared the MQDs brightness to four nearby single NRs spectra and plot $R(\theta) = I_{cross}/I_{single}$ for each fitted species (Supplementary Fig. 9a-c). At RT, we found the $R(\theta)$ of $X^B$ exciton and trion $X^{A-}$ increase and reach a maximum as the twist angle approaches ~20–28°, and then decrease at larger angles, whereas the $X^A$ exciton ratio shows no clear systematic dependence on twist angle. This might because under RT, Moiré related confinement is expected to be reduced due to thermally de-trapping of exciton, so twist-dependent optical contrast is less pronounced than at LT;[56–58] While $X^B$ and trion $X^{A-}$ are more responsive to local junction conditions such as stacking-dependent relaxation, localization strength, and carrier mobility.[7,9,59]

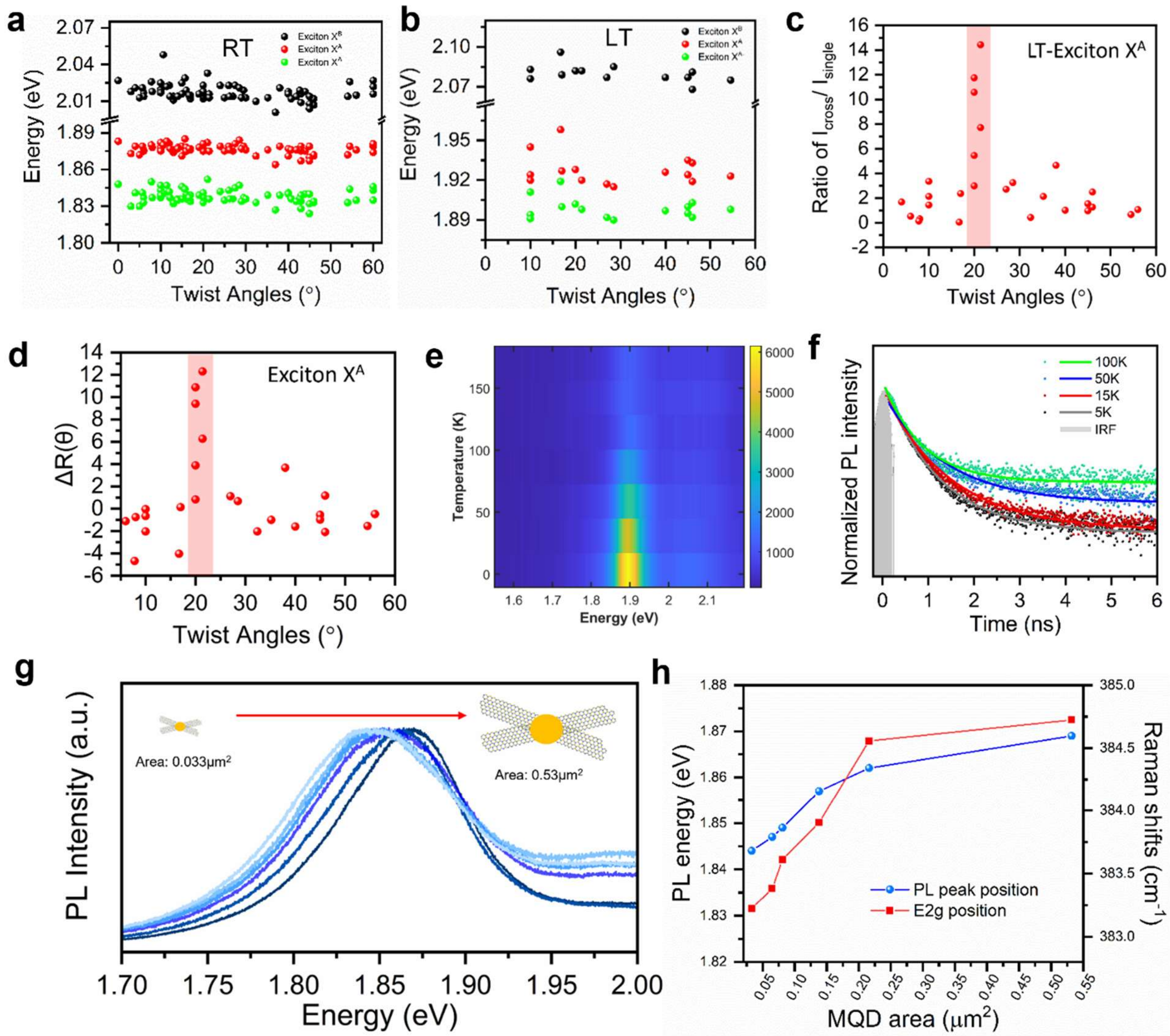

**Fig. 4: Twist angle-dependence exciton behavior of the MQDs a,b,** Fitted RT (**a**) and LT (**b**) PL energy position of exciton $X^A$ (red), $X^B$ (black) and trion $X^{A-}$ (green) as the function of twist angles (0° to 60°) on MQDs. **c,** Ratio of LT PL intensity on MQD and single layer ($R(\theta)$) for exciton $X^A$ as the function of twist angles. **d,** Difference of PL intensity Ratio ($\Delta R(\theta)$) for exciton $X^A$ on cross-section area between LT and RT as the function of twist angle. Pink columns in (**c**) and (**d**) indicate unique $R(\theta)$ and $\Delta R(\theta)$ values at $22\pm1°$ angle. **e,f,** Temperature dependent PL contour plot (**e**) from 3 to 170K and PL decay time (**f**) based in commensurate Moiré lattices at $22\pm1°$ MQD from 5 to 100 K, where $X^A$ decays fastest at 5K and becomes slower when temperatures increase into 100K. **g,** RT PL spectrum for 45° MQDs with different MQDs area sizes, the smallest area is 0.033 $\mu m^2$ and the largest is 0.53 $\mu m^2$, with (inset) atomic schematics showing MQDs area size change. **h,** Scattering plot of PL peak position and Raman $E_{2g}^1$ peak position with different MQDs areas for 45°.

To explore how the junction exciton landscape evolves with temperature, we carried out LT PL measurements at 3K on the same set of MQDs with different twist angles and performed Lorentzian fits to the $X^A$ and $X^B$ excitons and the trion $X^{A-}$ at the overlap region (Fig. 4b and Supplementary Fig. 8b). As expected, all three fitted resonances blue shift at low temperature relative to room temperature, consistent with the well-established temperature renormalization of $MoS_2$ exciton and trion energies arising from reduced electron-phonon interactions and thermal expansion effects.[60,61] However, the dominant effect is the twist-selective change in brightness of the MQDs relative to the adjacent monolayer NRs. Using $R(\theta) = \mathrm{I_{cross}}/\mathrm{I_{single}}$ and $\Delta R(\theta) = \mathrm{R_{LT}}(\theta) - \mathrm{R_{RT}}(\theta)$ (Fig. 4c and d and Supplementary Fig. 9d-g), we find that the $R(\theta)$ and $\Delta R(\theta)$ are strongly enhanced at 22° for exciton $X^A$ with $R_{\mathrm{LT}}$ and $\Delta R(\theta)$ reaching 14.4 and 12.5, respectively, indicating that cooling drives a disproportionate increase in exciton emission at the 22°commensurate angle MQDs when Moiré region is spatially restricted.

We further focus on this commensurate 22° MQDs and highlighted the temperature dependence by plotting the contour PL evolution from 3 to 170 K in the pseudo-color representation in Fig. 4e and overlapped plots (Supplementary Fig. 10a), where the MDQ emission is the strongest at cryogenic temperature and progressively quenches as temperature increases. Such behavior is consistent with a temperature-driven reduction of localization and spectral contrast, as widely observed in Moiré exciton systems where excitons thermally de-trap or become more delocalized at elevated temperature.[57,58] Complementary insight is provided by temperature-dependent Time-Resolved Photoluminescence (TRPL) in Fig. 4f, where the fitted $X^A$ exhibits the fastest decay at the lowest temperature and becomes progressively slower with increasing temperature. When considered together with the steady-state PL enhancement at low temperature, this trend suggests that the low-temperature exciton population is not simply more trapped, but also more radiatively efficient: localized exciton at the Moiré junction can recombine more faster through strong interaction, giving both higher PL intensity and a shorter effective lifetime. As temperature increases, excitons increasingly access delocalized, momentum-indirect, or dark-state reservoirs, which contribute less efficiently to photon emission and can introduce a longer-lived effective decay component.[13,62] In other words, the dominant exciton relaxation pathway evolves from a bright, localized radiative channel at low temperature to a more weakly emissive and spatially energetically redistributed population at higher temperature. Together with the large positive $R(\theta)$ and $\Delta R(\theta)$ at 22°, the contour and TRPL data support a distinct exciton relaxation pathway at this commensurate angle, consistent with prior reports that at 22° twisted bilayer $MoS_2$ hosts particularly pronounced Moiré-driven excitonic signatures resulted from shorter Moiré period, reduced effective mass, stronger interlayer interactions and periodic atomic reconstruction compared with other angles.[34,52,63]

In addition to twist angles, we demonstrated that the optical band gap of the MQDs depends strongly on junction size in Fig. 4g. For multiple MQDs prepared at the same nominal 45° twist angle, we find that smaller overlap areas (from down to around 0.033 $\mu m^2$) exhibit a systematic PL red shift of around 24 meV, compared with 0.53 $\mu m^2$ largest area, which is opposite to the blue shift expected from lateral confinement or disordered edge effect argument.[14,15] Instead, the correlated red shift in PL and the softening (red shift) of the in-plane $E_{2g}^1$ mode (Fig. 4h and Supplementary Fig. 10b) contribute to enhanced local tensile strain and edge-dominated lattice relaxation in smaller MQDs. Such structure reconstructions are known to drastically increase the depth of Moiré potential through induced strain, effectively lowering the $MoS_2$ excitonic transition energy and shifting Raman phonons.[14,64–66] Our results suggest that compared with 2D-2D Moiré platform, in smaller junctions, the increased edge-to-area ratio facilitates more pronounced reconstruction and indicating quantum dot-like behavior where the emission is governed by the depth of the local potential rather than the physical boundary.[67] This is also the first direct observation of a PL dependence on the size of the MQDs. These size-dependent excitonic landscape provides a natural route for MQDs area size variability and for amplifying twist-dependent bright-dark competition at 22° commensurate angle.

## Conclusion and outlook

In summary, different from conventional 2D-2D Moiré bilayers, our catalyst-free epitaxial synthesis of monolayer $MoS_2$ NRs on mica establishes a scalable platform for studying Moiré physics within laterally confined homostructures. By deterministic stacking of these 1D NRs, we create nanoscale junctions that function as quasi-0D MQDs, where the excitonic landscape diverges from extended 2D superlattices. This platform uncovers remarkable phenomena, including significant lattice relaxation from Moiré pattern and a sharp enhancement of exciton emission at the 22° commensurate angle, signaling a fundamental modification of exciton relaxation pathway when Moiré region is spatially restricted. Crucially, the NR junction introduces a highly accessible control parameter beyond twist: the overlap size. Reducing the MQDs areas from 0.53μm$^2$ to 0.033μm$^2$ produces a systematic 24 meV red-shift accompanied by $E_{2g}^1$ softening. This size effect confirms that edge-mediated lattice reconstruction drastically deepens the local Moiré potential, different from the exciton blue-shift at weakly confined width scales. Together, these results establish MQDs as a uniquely tunable platform where size, boundary-driven strain, and twist-angle converge to engineer exciton dynamics, which is difficult to realize in extended 2D-2D Moiré superlattice.

Looking forward, the extreme spatial confinement of MQDs poses a challenge of feasibility regarding diffraction-limited optical focusing and precise characterization at cryogenic temperatures. Future breakthroughs will likely necessitate the integration of near-field spectroscopy and magneto-optical techniques to probe these junctions with higher resolution. Specifically, applying external magnetic fields to the MQD system offers a promising trajectory to lift valley degeneracy and enable direct visualization of the dark-bright exciton competition within the Moiré landscape. Such investigations would clarify the impact of localized strain on spin-valley physics and evaluate the potential of MQDs as site-controlled, magneto-optically tunable quantum emitters for next-generation photonic circuits.

## Methods:

### CVD growth of MoS2 NRs

MoS2 NRs were grown by a catalyst-free one-step CVD method in a 1-inch diameter quartz tube under ambient pressure with high-purity argon as the carrier gas. Precursors of MoO3 (99.97%, Sigma-Aldrich) and NaCl (99%, Sigma-Aldrich) were mixed by mortar and pestle with weight ratio of 5:1, separately. Then 0.6mg of mixed MoO3/NaCl powders and 30mg of sulfur (99.98%, Sigma-Aldrich) powder were loaded in two alumina boats and placed on down- and up- stream separately. The freshly-cleaved mica (TaiYuan Fluorphlogopite Mica Company Ltd) was loaded on the top of the mixed MoO3/NaCl precursor boat. Before heating, 600sccm high flow argon gas was used to purge the tube for 30mins. The dwell temperatures for mixed MoO3/NaCl and sulfur powders are 750°C and 200°C, and the reaction time is 2mins. The furnace was opened and fast-cooled to room temperature when the growth was done. Monolayer MoS2 NRs would be formed on backside of the substrate (face-up) due to lowered density of MoO3 vapors during the growth.

## Fabrication of $MoS_2$ Moiré quantum dots

The first layer of $MoS_2$ NRs was formed from peeling off epitaxially oriented NRs on mica by PDMS film due to van der Waals force between $MoS_2$ and mica substrate. The PDMS film was initially attached on the sample surface of the mica substrate and pressed by cotton swipe for 5mins. During the peeling off process, the sample attached with PDMS was immersed into DI water to wet the mica surface, the uniform $MoS_2$ NRs can be lift off smoothly due to the hydrophobic and hydrophilic properties of 2D materials and mica substrate. Then, wet PDMS film with $MoS_2$ NRs was transferred to 285nm $SiO_2$/Si substrate and anneal under gradient temperatures from 60°C to 80°C and to 100°C for 5mins separately. The first layer of $MoS_2$ NRs can be formed on $SiO_2$/Si by peeling off PDMS quickly from as-transferred substrate. Similarly, the second layer of NRs on PDMS was stacked on the first layer of NR on $SiO_2$/Si, where the angle between the NRs can be controlled by the rotational transfer stage. After the same annealing process on hotplate, the PDMS film was peeled off, leaving behind the stacked NRs with controlled twist angles.

The stacked $MoS_2$ NRs were loaded into a CVD furnace and annealed again under argon atmosphere with 300°C for 1 hour. $MoS_2$ MQDs from stacked NRs with good interlayer coupling were formed after annealing.

## Scanning Electron Microscopy (SEM) measurements

Part of SEM images and EDS analysis were acquired through Microscope SEM JEOL IT800 with UHD detector and SHL mode. The other part of images and EDS analysis were from Microscope FEI Helios SEM/FIB equipped with EDS detectors.

## Atomic Force Microscopy (AFM) measurements

The topography and phase images and NRs height information were carried out from Microscope: NX20 AFM under tapping mode with 0.2Hz scanning rate.

## Room Temperature (RT) Raman and Photoluminescence (PL) measurements

RT Raman and RT PL were collected from Renishaw-inVia Raman Microscope with a 532 nm laser and 1800 $mm^{-1}$ grating.

## Low Temperature (LT) Photoluminescence (PL) and Time-resolved Photoluminescence (TRPL) and laser-scanning confocal PL imaging

The PL measurements were performed on a home-built spectroscopy system (Horiba iHR 320 monochromator, Syncerity EM-cooled CCD) with the optical path connected to a liquid-helium recirculating optical cryostat (Montana Instruments CryoAdvance50 with Cryo-Optic module). A vacuum-compatible 100X, NA 0.90 objective (Zeiss LD EC Epiplan-Neofluar 100x/0.90 DIC M27) was mounted inside the cryostat. A CW laser centered at 532 nm with a maximum output power of 36 mW was used as excitation. All the PL spectra are reproducible in repeated measurements at the same spot on the sample.

All TCSPC measurements were performed based on the MicroTime100 microscope (PicoQuant) coupled with the TimeHarp 260 time-counting module (PicoQuant), with an optical path to Montana Instrument CryoAdvance50 with a Cryo-Optic module, the same as the steady-state spectroscopy measurements. 509 nm 40 MHz pulse laser (PicoQuant LDH-D-C-520) was used to excite the sample. The emission signal was selected by hard-coated bandpass filters center at specific peak energies, then collected by single-photon avalanche photodiodes(SPAD) (Excelitas SPCM-AQRH-14-TR).

The laser-scanning confocal PL imaging was achieved on a home-built confocal microscopy system using a voltage-controlled scanning mirror and a 4f lens configuration. The sample was excited by a CW 488nm laser with 5mW optical power, which was then focused onto the sample in cryostat through an objective lens (50×, 0.65 NA, Mitutoyo). PL spectrum at each scanned pixel was collected by a spectrometer (Kymera 328i, Andor) coupled with a CCD camera (iDus 416, Andor).

## Second Harmonic Generation microscopy (SHG) measurements

For the SHG measurements, a Light Conversion FLINT femtosecond oscillator was used to drive an APE Levante IR optical parametric oscillator (OPO) coupled with an APE HarmoniXX SHG/THG module. This system delivered 120 fs pulses at a 76 MHz repetition rate with a fundamental wavelength of 800 nm. The fundamental beam was set to linear polarization with a Glan–Taylor polarizer. The beam was focused onto the sample at normal incidence using a 40× objective (f = 5 mm, NAeff = 0.6), with a beam diameter of 6 mm, yielding a spot size of approximately 0.5 µm. The incident power on the stacked $MoS_2$ NRs was maintained at 3 mW. In the home-built reflection-geometry SHG setup, the rotational mount was integrated with a motorized scanning stage to enable spatially resolved SHG imaging, from which the intensity count was calculated. The reflected SHG signal was isolated via a 785 nm short-pass filter and a 400 nm band-pass filter, and passed through a thin-film analyzer aligned parallel to the input polarization. The signal was detected using a Hamamatsu C11202-050 single-photon counter.

### Transmission Electron Microscopy (TEM) measurements

The samples were prepared by spinning coating the PMMA on top of $MoS_2/SiO_2$ substrate, followed by immersing samples into the KOH to etch the $SiO_2$ off. Floated PMMA/$MoS_2$ film was further scooped up and transferred to water to wash the KOH away and then scooped up by TEM grid and immerse into the acetone to remove the PMMA. Sample examined using an aberration-corrected FEI Titan Themis operated at 80 kV. For atomic-resolution imaging of $MoS_2$, a convergence semi-angle of 25.2 mrad was employed, providing sufficient spatial resolution to resolve both the atomic lattice and Moiré superlattice features. The abTEM framework is simulated from: https://abtem.readthedocs.io/en/latest/intro.html

**Acknowledgments:** We thank Z. Huang and X. Ma for useful discussions on the magneto optical spectroscopy measurements. We thank A. Jade for part of twist angle verification.

**Funding:** X.S. and P.M.A. acknowledge the grant from DOD- AFRL UES project (FA8650-22-D-5814). T.A.M.R.S and H.Z. acknowledge support from the Welch Foundation C-2311. W.W. and S.H. acknowledge the funding support of National Science Foundation (Grant Nos. ECCS-2246564, ECCS-1943895, and ECCS-2230400), the Office of Naval Research (ONR, Grant No. N000142512387), and the Robert A. Welch Foundation (Grant No. C-2144). The author L.D. acknowledges the fellowship grant in Rice Mehta Engineering Scholar Program- 2024 for providing an opportunity to work at Rice University USA.

**Author Contributions:** P.M.A., X.S. conceived the idea, designed the experiments, analyzed the data and cowrote the manuscript. H.Z., R.N. and A.D.M. performed LT PL and mapping with X.S; W.W. and S.H. helped with LF Raman, LT PL and TRPL measurements; C.W. and D.P. helped with sample growth with X.S.; M.A. and T.P. performed TEM and STEM and analysis. T.A.M.R. and H.Z. performed SHG measurement and fitting; D.P., L.D. and Y.Z. helped with sample transfer and stacking with X.S.; C.W., Y.Z., L.V.L. and L.S. performed the twist angle verification. X.Z. helped with SEM EDAX measurements. B.K.G., H.Z., Z.K.N., X.Z., A.P., J.L., R.V. and A.D.M. reviewed and edited manuscript. All authors contributed to the manuscript.

**Competing interests:** Authors declare that they have no competing interests.

**Data and materials availability:** All data are available in the main text or the supplementary materials.